# Word-of-mouth and dynamical inhomogeneous markets:

# Efficiency measure and optimal sampling policies for the pre-launch stage

ELENA AGLIARI

*Theoretische Polymerphysik, Universität Freiburg, Hermann-Herder-Str. 3*

*D-79104 Freiburg, Germany*

RAFFAELLA BURIONI[#] AND DAVIDE CASSI

*Dipartimento di Fisica, Università degli Studi di Parma and INFN Gr*uppo Collegato di Parma

Via G.P. Usberti 7/A, 43100 Parma, Italy

AND

FRANCO MARIA NERI

*Department of Plant Sciences, University of Cambridge,*

*Downing Street, Cambridge, CB2 3EA, UK Cambridge*

*ABSTRACT*

An important assumption lying behind innovation diffusion models and word-of-mouth processes is that of homogeneous mixing: at any time, the individuals making up the market are uniformly distributed in space. When the geographical parameters of the market, such as its area extension, become important, the movement of individuals must be explicitly taken into account. The authors introduce a model for a "micro-level" process for the diffusion of an innovative product, based on a word-of-mouth mechanism, and they explicitly consider the inhomogeneity of markets and the spatial extent of the geographical region where the process takes place. This results in an unexpected behaviour of macro (aggregate) level measurable quantities. The authors study the particular case of the pre-launch stage, where a product is first presented to the market through free sample distribution. The first triers of the samples then inform the other potential customers via word-of-mouth; additional advertising is absent. The authors find an unexpected general failure of the word-of-mouth mechanism for high market densities and they obtain quantitative results for the optimal sampling policy. By introducing a threshold to discriminate between individuals who will purchase and those who will not purchase according to their individual goodwill, they calculate the length of the pre-launch campaign and the final goodwill as a function of the firm's expenditure. These results are applied to a set of major US urban areas.

*Keywords*: Diffusion models; simulations; random walks; word-of-mouth; pre-launch strategies.

## 1. Introduction

The firm that introduces an innovative product needs an optimal advertising strategy in order to improve market's awareness and goodwill. Marketers often handle this problem with the aid of diffusion models and attempt to describe the spread of a new product innovation among a set of potential adopters as a "viral" contagion (see, for example, the book by Rogers (1983)).

In the framework of diffusion models, it is commonly believed (Bass, 1969; Rogers, 1983; Midgely, 1988; Dockner & Jorgensen, 1988; Goldenberg *et al.*, 2002), on the basis of behavioral theories, that there are two main types of sources which convey information to potential consumers: 1) firm-originated advertising, which mainly affects the so-called "innovators", and 2) word-of-mouth communication from individuals who are aware of the innovation to imitators. Hence, diffusion is started by advertising and goes on through word-of-mouth (henceforth, WOM). As an alternative to advertising, the distribution of free samples (Jain *et al.*, 1995; McGuinness *et al.*, 1992; Heiman *et al.*, 2001) has been considered as an effective means for introducing and promoting a new product: in these models, diffusion is started by free sampling *and* advertising and goes on through WOM .

Though not within the direct control of the firm, it is well-established (Glaister, 1974) that WOM plays an important role in knowledge diffusion: we tend to be suspicious of advertising claims from the seller, while information coming from an "impartial" user who has experience of the good in question is often more influential. Knowledge diffusion between individuals strongly relies on direct interaction and face-to-face contacts (Cowan & Jonard, 2004). Herr *et al.*, (1991) give evidence that information transferred by a face-to-face process is more "vivid", and consequently more persuasive, than firm-originated information. A plenty of studies, both theoretical and empirical, describe the spread of information about a new product among a set of potential consumers taking into account this mechanism (see for example Horsky & Simon, (1983), Monahan, (1984), Dockner & Jorgensen, (1988), Vettas, (1997), Banerjee & Fudenberg, (2004), East *et al.*, (2007), and a review on stochastic models about consumer response by Hauser & Wisniewski, (1982)).

The usual assumption lying behind diffusion models in marketing is that of *homogeneous mixing*: it consists in considering all the individuals of the market as homogeneously distributed in space. Our aim is to obtain a deeper insight into WOM processes by relaxing this assumption, and introducing spatial inhomogeneity in population movement and distribution. This leads to unexpected results in optimal sampling policy.

We consider a model for the "micro-level" WOM process, incorporating three main features aimed at defining a more realistic representation of diffusion processes: i) we explicitly take into account the motion of the individuals who make up the market; ii) we introduce heterogeneity among individuals (each of them can have a different individual goodwill and a different position), iii) we introduce a decay of goodwill as the information about the product passes from an individual to another.

We will focus our attention on a monopolistic market, with a given geographical area, in the pre-launch phase of a new product. During this stage, a free-sample distribution is performed without any auxiliary advertising from the firm. This way, the firm makes some people know about its new product (Lehmann & Esteban-Bravo, (2006)). Those sources (the first triers of the product) will then inform unaware people, who will in turn be able to spread information through WOM. We assume that during information spreading the individual goodwill about the new product spontaneously depreciates when the information passes from one individual to another. When all the market has been reached by the information, the product is finally released, and further advertising can be possibly performed (Sonnemans P.J.M. *et al.*, 2003).

Our model yields three main results; two of them are quantitative. First, we can forecast the length of the pre-launch stage, defined as the time required for the information about the new product to reach all the population. This time is a non-trivial function of the total population, the geographical area of the market and the number of free samples distributed (hence the firm expenditure). Second, we can calculate the final amount of goodwill as a function of the same variables, and find an optimal sampling policy: that is, how many samples to distribute to improve consistently the number of potential buyers. The third result is qualitative, and is related to the emergence of a local minimum of the final goodwill as a function of the density of the market area. This minimum corresponds to a loss of efficiency of the WOM process and is due exclusively to geographical reasons: it should appear for particular values of the market density for any launch campaign that relies heavily on WOM effects.

The main focus of our work is analytical and numerical, and is intended to highlight possible strategies for the seller before introducing the product on the chosen market. The possibility of pre-launch measurements of individual characteristics (based on established marketing science procedures) points to potential applications of the model for the segmentation of the potential adopter population.

## 2. Homogeneous mixing, diffusion models and random walks

The hypothesis of *homogeneous mixing* (Gleister, 1974) states that every member of the population is equally likely to interact with each other, the chances of contact between any two individuals being the same at every instant (akin to the *perfect structure* of Frenzen & Nakamoto, (1993): every person shares a social relation with every other person). Therefore, as an example, the number of potential adopters of a product at a given time depends only on *global* quantities, such as the *total* population and the *total* number of current users at that time. However, in fact, each potential adopter can only learn about the new product from the users he sees *at that time*, and the number of such users may vary from point to point in space. What matters is the *contact structure* of the knowledge network in that instant.

The implicit idea behind homogeneous mixing is that the diffusion of individuals happens at a much more rapid time scale than every other process (in particular, information transfer): when this happens, every individual is able to meet every other individual before the information transfer has been concluded, therefore averaging over the whole population. This is not the case for most real systems, where information passing takes place *while* the individuals move in space. Information passing is much more rapid than physical movement: face-to-face contacts are often instantaneous, the pertinent information being passed in only a few seconds from one person to another. Thus, both processes occur on the same "time-scale", that is, each instant corresponds to a spatial distribution of people and to a consequent information transfer according to the set of people "in contact". This explains why the spatial distribution of individuals and the consequent *market inhomogeneity* have to be taken into account for a realistic description (Karmeshul & Goswamil, 2001).

Hence, the way the population moves is fundamental for advertising based on word-of-mouth. A better insight can be achieved from the way diffusion processes are treated in physics and chemistry. The basic model for diffusion processes is provided by Random Walks (RW) (Weiss, 1994; Burioni & Cassi, 2005), which are a simple stochastic process, possibly represented as the path described by a walker starting somewhere in space an taking steps in a random direction.

The theory of random walks has been extensively studied by means of theoretical models, experiments, and computer simulations and is widely applied in several fields, ranging from physics to economics to biology. The reason is that the behavior of a large class of natural and social systems can be cast into the form of a random walk. Especially in physics and mathematics, random walks have been widely studied both on regular structures such as lattices and on generic graphs.

In particular, the *simple random walk* on a discrete structure can be formalized as follows. Assuming the time $t$ to be discrete, we define at each time step $t$ the jumping probability $p_{ij}$ from site $i$ to an adjacent site $j$:

$$p_{ij}(t) = \alpha_i^{-1},$$

where $\alpha_i$ represents the degree (number of adjacent sites) of $i$. Notice that this is the simplest case we can consider: the jumping probabilities are isotropic at each point and they do not depend on time; moreover, the walker is forced to jump at every step.

Works dealing with the theory of RW do not consider only problems involving just a single ($N = 1$) random walker. There are interesting diffusion problems which imply many random walkers, and for which the diffusion process cannot be described by averaging over the properties of a single walker (Yuste & Acedo, 1999).

We introduce a model that takes into account the point of view described above: individuals are represented by random walkers diffusing on a two-dimensional lattice; they can interact (that is, pass product information to each other and change each other's goodwill) only if they are on adjacent sites. The resulting society is therefore dynamic, and contacts between individuals are instantaneous. Hence, at each instant the number of aware individuals depends both on the number of the current users and on their reciprocal positions.

## 3. The model

A first version of this model was studied in the context of social networks; (Agliari *et al.*, 2006). Starting from a population, we define the "market" as the fraction of the population that is made up of identical individuals who are potential buyers. We also assume that no potential buyer can become aware of the new product spontaneously or by firm advertising. The market is composed of $N$ potential purchasers (henceforth, individuals) that move in a well-defined geographical space of area $L^2$; this can be a whole country or, more realistically, an urban area. In our model, individuals are represented by random walkers moving on a square $L \times L$ lattice with periodic boundary conditions[1]; $\rho = N/L^2$ is the population global density.

---

[1]It is a common practice in numerical simulations to adopt periodic boundary conditions, though they often correspond to non-realistic situations: the reason is that they speed up considerably the simulation time. We have checked that, by

Before the simulation starts, the firm introduces a given amount of free samples in the market. We assume that the product launched is of high quality in such a way that the $N_s$ individuals (henceforth, "sources") who first test the samples will surely become purchasers of the product at the end of the campaign; hence, they are assumed to have perfect information and goodwill about the product. We also suppose that $N_s$ is proportional to the total number of samples distributed, and that the latter is proportional to the total expenditure of the firm in the absence of advertising (that is, we neglect any lump set-up cost). The sample distribution ends at time $t = 0$. At this time, the individuals are distributed on the lattice with a given configuration; in our simulation, we choose the most convenient, that is a random distribution. There can be more individuals on the same site. At $t = 0$, the $N$ individuals ($N_s$ informed, $N - N_s$ unaware) start moving on the lattice. At each instant $t > 0$ each individual jumps randomly to one of the four adjacent sites.

Each individual $j$ carries a number $G_j$, $0 \leq G_j \leq 1$, representing goodwill; goodwill is gained by receiving information about the product, either from the direct test of a sample or by WOM. An individual is called "informed" if $G_j > 0$ and "unaware" if $G_j = 0$. At $t = 0$ the $N_s$ sources are informed and carry goodwill 1; the other $N - N_s$ individuals are unaware. The aim of the dynamics is to diffuse information from the sources to the whole market.

Two individuals are "in contact" if they are either on the same site or on adjacent sites. Interaction between two individuals in contact takes place when one of them is informed and the other still unaware of the new product. By "interaction", we mean information passing from the informed individual to the unaware one. The simulation stops at the time when all the individuals have become informed; this time is a stochastic variable whose average we call the Market-Awareness Time $\tau$. We call $n(t)$ the total number of informed individuals at time $t$ ($n(0) = N_s$; $n(\tau) = N$)).

---

adopting more realistic conditions (e.g., a reflecting boundary), the numerical results we present are not substantially

The effects of inhomogeneity have seldom been object of study, with a few exceptions (for example, Putsis *et al*., (1997); Allaway *et al*, (1994); Van Den Bulte & Joshi (2007)). We also remark that the delay between the sample distribution and the time the information reaches every individual, which is central in our model, could remind lag models (see Feichtinger *et al*, (1994) for a review). The latter suppose there is a time lag between the advertising expenditure (in our case, the sample distribution) and the increase in the stock of goodwill. This is a different hypothesis from ours: we are not forced to give *ab initio* the functional form of the lag, as in those models, since this comes as a consequence of the particular kind of movement we have chosen.

We suppose that by passing from one aware individual to another the goodwill naturally loses strength: there can be loss of precision in the information about the new product, or loss of vividness (Herr *et al*, 1991). We simulate this loss by multiplying the goodwill of the aware agent by a decay number ($0 < z < 1$) when passing information to the unaware. It is of course very difficult to guess the exact form of this decay function, be it time-dependent, individual-dependent, etc. We choose here a particular form: $z$ is a constant[2]; the exact value of this constant for a particular market could be extracted, for example, by means of market surveys after the pre-launch campaign.

This way, we drop another often-exploited hypothesis in diffusion models, that of *Perfect cooperation* (Frenzen & Nakamoto, 1993): it implies that all people will transmit the information they acquire perfectly, that is, without any correction or revision, so that there cannot be misrepresentation of quality of the product (Rogerson, 1983). This hypothesis has been carefully reconsidered: it has been supposed that the essence of the news can spread being, at least partially, revised due to idiosyncratic occurrences or private interests (Frenzen & Nakamoto, 1993). Chatterjee & Eliashberg, (1990) suggested a micromodeling approach where the opinion is strongly individual due to a personal

---

changed.

[2]The results do not substantially vary if we choose, for example, a complementary form, where $z$ is a random variable extracted from a uniform distribution in the interval $[0,1]$.

evaluation, and the adoption times can be different due to heterogeneity in the population. Easingwood *et al*, (1983) state that assuming that the efficiency of information transmission between individuals remains uniform over the entire span of the process is quite restrictive; they suggest a model where efficiency is time dependent (nonuniformity in time). More in general, the decay of the goodwill stock is attributed in most models to forgetting phenomena (starting from Nerlove & Arrow, (1962)).

However, apart from a purely time-dependent decay, the efficiency of information spreading (and the final amount of goodwill) is also affected by the number of passages among different individuals. Each time the piece of information is transferred form one individual to another, it is revised, due to loss of vividness and precision. In general, the passing-decay effects and the forgetting effects would overlap; in this work we only consider the pre-launch phase, hence we assume that the time framework is short enough for no forgetting to occur (as in Mahajan *et al*, (1984)). Consequently, we expect that the more passages the information has experienced, the more different from its original form and strength it has become. Hence, two different individuals, though both informed, can own different goodwill; this is very important at the end of the pre-launch stage, when every individual has received the information and the total goodwill must be evaluated.

As a result of our model, the goodwill carried by an individual depends only on the number of passages $l$: its value is $G_i = z^l$. We say that an informed individual belongs to generation $l$ when it has received information after $l$ passages from a source. We call $n(l,t)$ the number of individuals belonging to the $l$-th generation at time $t$: $n(t) = \sum_{l=0}^{t} n(l,t)$.

At each instant $t$ we define the total goodwill

$$\mathcal{G}(z,t) = \sum_{l=0}^{t} n(l,t) z^l; \tag{1}$$

We are interested in particular in the final goodwill

$$\mathcal{G}(z) = \mathcal{G}(z,\tau).$$

and in its average value per individual, $\mathcal{G}_{av}(z) = \mathcal{G}(z)/N$.

We also quantify the individual awareness so that the consequent choice can be derived. In fact, in the first part WOM makes individuals aware of the new product and, in our model, individuals are consequently assigned an awareness status according to the amount of goodwill they hold. This distribution reflects the initial expenditures made by the firm. Moreover, based on this distribution and other market parameters (as we will see, decay constant and threshold), we determine the number of those people who intend to purchase.

The final goodwill is the quantity from which the final number of purchasers is usually computed. Since our model also deals with *individual goodwill*, we suggest a different way to calculate the final number of purchasers, namely, by means of a threshold.

Since agents who achieve a high level of goodwill are more likely to become purchasers, a threshold $\xi$ ($0 < \xi < 1$) is introduced in order to discriminate amongst buyers and non-buyers. Consumers may be skeptical about products that they have heard about only approximatively or vaguely; hence, we state that an individual $j$ adopts the new product whenever his goodwill $G_j$ is large with respect to price and risk issues. We quantify this statement by assuming that individual $j$ will purchase whenever $G_j > \xi$, where $\xi$ is a threshold value, suitably introduced and representing the price and risk "hurdles" to be overcome. We then calculate the percentage of individuals who, at time $t = \tau$, would buy the product according to the threshold introduced.

The independent variables we will use in the analysis are: the market size $N$, the number of sources $N_S$ (i.e., the firm initial expenditure), and the geographical market area $L^2$. Two more independent variables are the decay parameter $z$ and the threshold $\xi$; these variables are intrinsic and can be recovered only through a post-process study of the market behavior.

The quantities we are especially interested in are:

• The market awareness time $\tau$ : at time $t = \tau$ each individual in the market has been informed, is aware about the new product and has a certain degree of individual goodwill about it;

• The final average amount of goodwill $\mathcal{G}_{av}$ .

• The percentage of individuals who at time $t = \tau$ would buy the product according to the threshold introduced.

Our model is theoretical and based on a number of simplifying assumptions that we will underline and discuss in the following; further empirical analysis would clarify how to improve the model. The major assumptions we have done are:

• We choose the square lattice as substrate throughout which the diffusion and the interaction of agents occur. Other topologies have been considered in other contexts (Agliari *et al.*, 2007, Agliari *et al.*, 2007) giving rise to analogous results. The foundamental point in all these cases is that the "contact network" (i.e., the network describing who is interacting with whom) is continuously revised according to the instantaneous position of each agent.

• The market has a given spatial density, depending on $N$ and $L$ : the global density is fixed and equal to $\rho = N/L^2$ , while the local density (relevant to a given subregion) depends on time, due to the motion of agents. Moreover, since both $N$ and $L$ are finite, sooner or later the information will reach each individual.

• The model does not consider the possibility of forgetting or updating, or the existence of external advertising. We also assume that there is no negative WOM, that is, the initial goodwill spreading from the sources is positive and equal to one (this is a realistic assumption for high-quality products).

• Since we assume that advertising from the firm is absent and the only decision variable concerns the product sampling, the only firm expenditures are due to free samples. Hence, the final profit reached by the firm will be compared with initial cost in sampling.

• Agents who achieve a high level of goodwill are more likely to become purchasers. Hence, we model satisfaction by using a threshold-crossing formulation whereby, if the goodwill of a given agent exceeds a threshold level, that agent is a buyer.

• As previously assumed, we ignore the possibility that there is a segment of the population which, though it may purchase the product, does not participate in the WOM process.

## 4. Results

Though our model is relatively easy to formulate, its analytic solution is quite complicated and it shall be considered in the Appendix (in particular, we will focus on the high-dilution regime). On the other hand, this model is suitable to treatment by means of Monte Carlo (MC) simulations. Indeed, this system displays a crucial stochastic element, since at each time $t$ each individual moves towards a randomly selected adjacent site. Hence, MC simulations allow us to deal with the problem in a probabilistic fashion, so that good results (i.e., with the desired accuracy) can efficiently be obtained. This numerical technique displays an exceedingly wide range of different applications in economics (Glasserman *et al*, 2000; Fu & Hu, 1999), physics (Binder & Heermann, 1988), biology, and even problems of traffic flow (Herty *et al*, 2005).

Our simulations start at the time when the product sampling has just been performed, all the parameters presented in the previous section being fixed. We analyse several scenarios by changing number of sources and density separately. We remark that for most numerical simulations we will use values for the parameters $N$, $L$ and $N_s$ that are quite small with respect to real markets. This is done for the sake of computability; the point is that, thanks to the fitting procedure, the behavior of the system for larger parameters can be *extrapolated* from the smaller-parameter simulations.

## 5. Sigmoidal curves and market awareness time

We first focus our attention on the shape of the growth curve $n(t)$, representing the total numer of informed individuals as a function of time, and on the rate of diffusion (see for example how Hauser & Wiesniski (1982) and Chatterjee & Eliashberg (1990) treat this problem, also in relation with sales).

FIGURE 1: Sigmoidal growth curve $n(t)$ for N=32, $N_s = 1$, L=512.

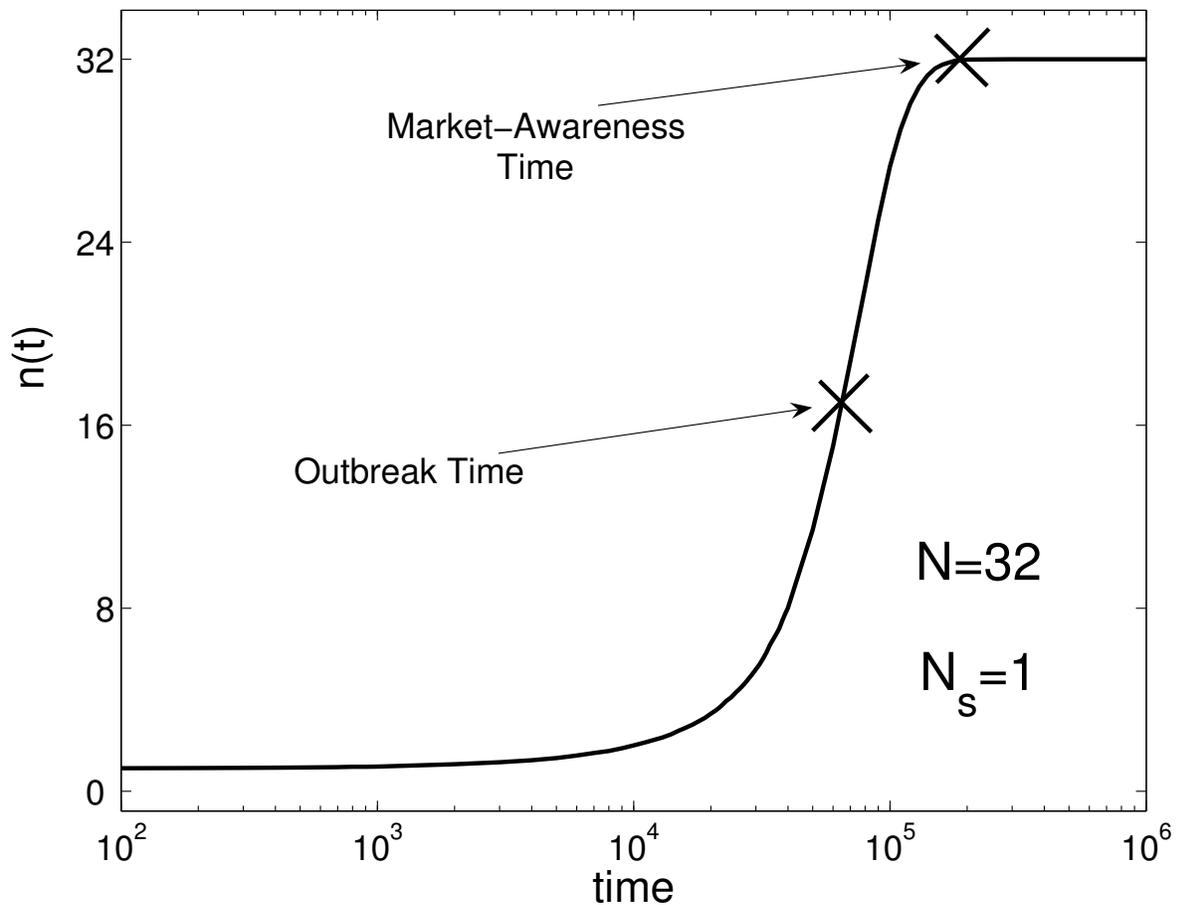

Figure 1 shows $n(t)$ for a market with parameters $N = 32$, $N_s = 1$, $L = 512$. A result of our simulations is that $n(t)$ follows a sigmoidal law. The saturation value is $n(\tau) = N$, where $\tau$ is the Market-Awareness Time, when the information has reached the whole population. This time gives the seller an important indication: how fast the piece of information spreads among the population or, otherwise stated, how long it takes the whole market to become aware. According to our model, no improvement in the final amount of goodwill can be attained by waiting further after $\tau$. Hence, time $\tau$ should be considered as the maximum time interval to wait before releasing the product, since by waiting further forgetting phenomena may come into play.

Another important quantity, the Outbreak time $t_{out}$, is defined as the position of the flex of $n(t)$, that is, the instant when the rate of diffusion (or, in other words, the spreading efficiency) is maximum. Its value is in general about $\tau/2$; hence, its dependence on market parameters can be deduced from that of $\tau$.

The only parameter the firm can tune is the number of sources: if the firm initially makes a larger number of free samples available, the Market-Awareness Time will decrease. The fitting of the data for $\tau$ obtained by numerical simulations is in agreement with the expression derived analytically in the appendix, namely:

$$\tau = C \frac{L^2 \log L}{N} \log \frac{N(N-N_s)}{N_s}. \qquad (2)$$

The fitting procedure allows estimating the value of the constant $C$ treated as a free parameter: using the least-squares method the fitting yields $C = 0.30 \pm 0.02$. The functional dependence of $t_{out}$ is the same as for $\tau$ (Equation 2), with the coefficient $C$ halved.

As an example, we assume a pre-launch campaign organized on an urban-area basis: the distribution of free samples (hence, the parameter $N_s$) is different from town to town. Equation 2 shows that, given two different urban areas, this time depends on their area $L^2$, total population $N$ and initial expenditure $N_s$.

Let us consider, e.g., some of the principal US urbanized areas (see the data released by the US

Census Bureau on the web site "USA Urbanized Areas 2002"). The area is equal to the parameter $L^2$ of our model. Let us suppose that the overall market for our product consists of exactly one tenth of the total population. We use generic time units: in order to set a timescale, we take as a reference the time $\tau = 1000$ required to inform the whole city of Philadelphia with an initial expenditure $N_s = 1000$. Then, given the population and urban area of Philadelphia, the $\tau$ for all the other cities (for three different values of $N_s$) can be calculated with respect to that of Philadelphia knowing their values of $L^2$ and $N$, as shown in Table 1.

TABLE 1: Expected Market-Awareness Times for some of the US major urban areas.

| | Population | Area | $\tau(N_s = 100)$ | $\tau(N_s = 1000)$ | $\tau(N_s = 10000)$ |
|---|---|---|---|---|---|
| New York | 17799861 | 3352.60 | 696 | 631 | 566 |
| LosAngeles | 11789487 | 1667.93 | 483 | 436 | 390 |
| Chicago | 8307904 | 2122.81 | 860 | 774 | 689 |
| Philadelphia | 5149079 | 1799 | 1115 | 1000 | 884 |
| Miami | 4919036 | 1116.09 | 698 | 625 | 553 |
| Boston | 4032484 | 1736.18 | 1341 | 1199 | 1056 |
| Detroit | 3903377 | 1261.45 | 982 | 878 | 773 |
| Atlanta | 3499840 | 1962. | 1738 | 1551 | 1363 |
| SanFrancisco | 3228605 | 526 | 457 | 407 | 358 |
| Phoenix | 2907049 | 799.01 | 787 | 701 | 614 |
| Seattle | 2712205 | 953 | 1012 | 901 | 788 |
| San Jose | 1538312 | 260.11 | 419 | 320 | 406 |

| Columbus | 1133193 | 387.71 | 870 | 765 | 658 |
|----------|---------|--------|-----|-----|-----|

Interestingly, since the variable $N_s$ appears under a logarithm, its effect on $\tau$ can never grow too large (unless $N_s$ is of the same order of magnitude of $N$, which we suppose to be too expensive an option for the firm). Typically, an increase by a factor 10 of $N_s$ leads to a decrease of $\tau$ of about $10-12\%$. Hence, a significant decrease of $\tau$ requires an exponential increase of the initial expenditure and should be considered only when a very quick pre-launch phase is strictly necessary.

## 6. Final amount of goowill

We now analyze $n(l,t)$, the time evolution of the population belonging to the $l$-th generation. Each population evolves in time according to its own sigmoidal law, tending to its saturation value $n(l,\tau)$. In particular, low generations, which are expected to receive less altered information, are also the ones that saturate earlier. The saturation values $n(l,\tau)$ follow an asymmetrical bell-shaped distribution, peaked at $l_{peak}$ and with width $\sigma$; both parameters depend on $N$, $L$ and $N_S$. The analytical expression for $n(l,\tau)$ can be calculated only in a low-density approximation, as shown in the appendix.

The curves $n(l,\tau)$ are of great importance for the seller, since they provide an efficient description of the population goodwill at time $\tau$. In fact, as a result of WOM process, the population is divided into several generations, so that people belonging to the same generation $l$ hold a piece of information which has undergone the same number of passages $l$. Hence, one expects that the lower the generation, the more persuasive the message conveyed and the larger the goodwill. In particular, the total amount of final goodwill will be higher if the curve is peaked on lower values of the generation $l$, and lower if the curve is peaked on higher values of $n$. It is therefore extremely interesting to see how the curves $n(l,\tau)$ change by varying the market parameters $N$, $L$ and $N_s$.

Let us consider a market characterized by a fixed decay constant $\bar{z}$. Figure 2 shows how the distribution of individuals on generations changes by keeping constant the market $N$ and varying the market area $L^2$, with a fixed number of sources (in this case $N_s = 1$). The first two frames on the left show the distribution for several values of $L$; the frame on the right shows the variation of the final goodwill for the whole range of $L$. The best results correspond to large densities, since in that case the piece of information has to undergo just a few passages, so that, on the average, it is not significantly altered from the original. It turns out that the distribution is initially (for low values of $L$) very narrow and peaked on small values of $l$; the corresponding value of the average final goodwill is high. By reducing the market density, one expects that the piece of information has to experience a greater number of passages and then to get more and more altered. By increasing $L$, the distribution flattens, and its peak shifts to higher $l$: correspondingly, the final goodwill decreases. This behavior goes on up to a definite value of $\tilde{L}$, that depends on $N$ and (more weakly) on $N_s$.

For the case chosen here, $\tilde{L} = 64$. For this value of the market size we obtain a maximally spread curve, and the minimum value of the final goodwill. For $L > \tilde{L}$ (middle frame), the curves begin to shift back and to sharpen, and the final goodwill increases again.

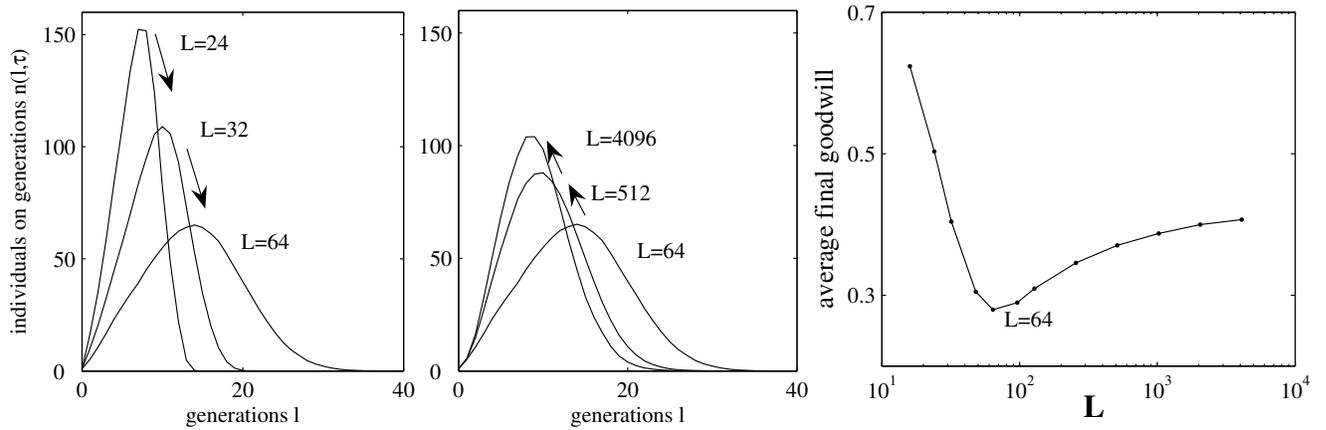

Summing up, there is an *optimal* area for which the distribution $n(l, \tau)$ is rightmost extremal, and the value of the final goodwill is minimum. Analogously (not shown here), once $L$ is fixed, there exists a minimum for $N = \tilde{N}$, where $\tilde{L}$ and $\tilde{N}$ depend on $N$ and $L$, respectively. Figure 3 shows how the depth of this minimum depends on the decay parameter $z$.

It must be underlined that the existence of an extemal point does not depend on the particular topology chosen (here a square torus), but has been proved also for $d$-dimensional Euclidean lattices and fractals (Agliari *et al.*, 2007, Agliari *et al.*, 2007).

FIGURE 3: Final average amount of goodwill; $N = 1024$, $N_s = 1$.

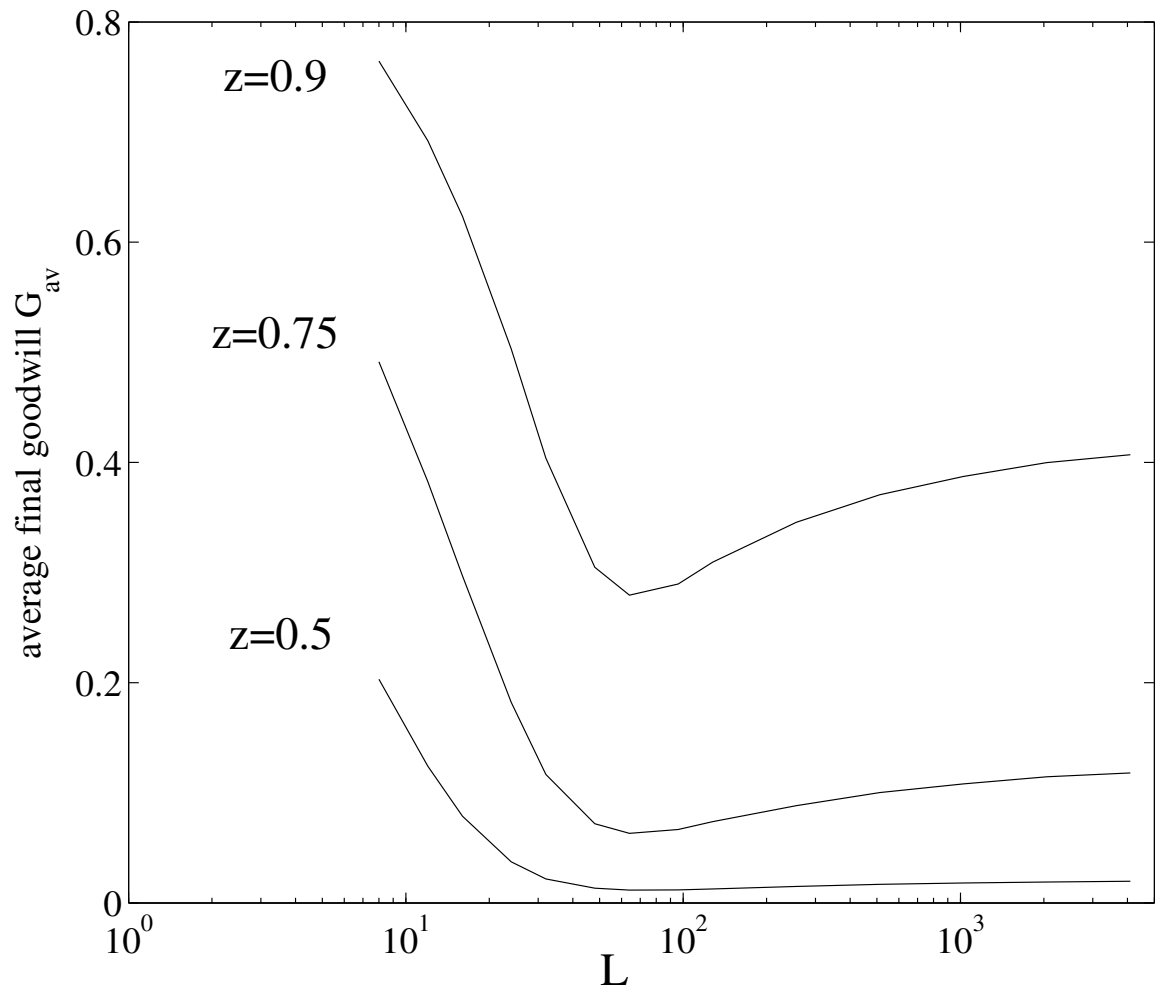

FIGURE 4: Distribution of individuals in generations and final average goodwill for different $N_s$.

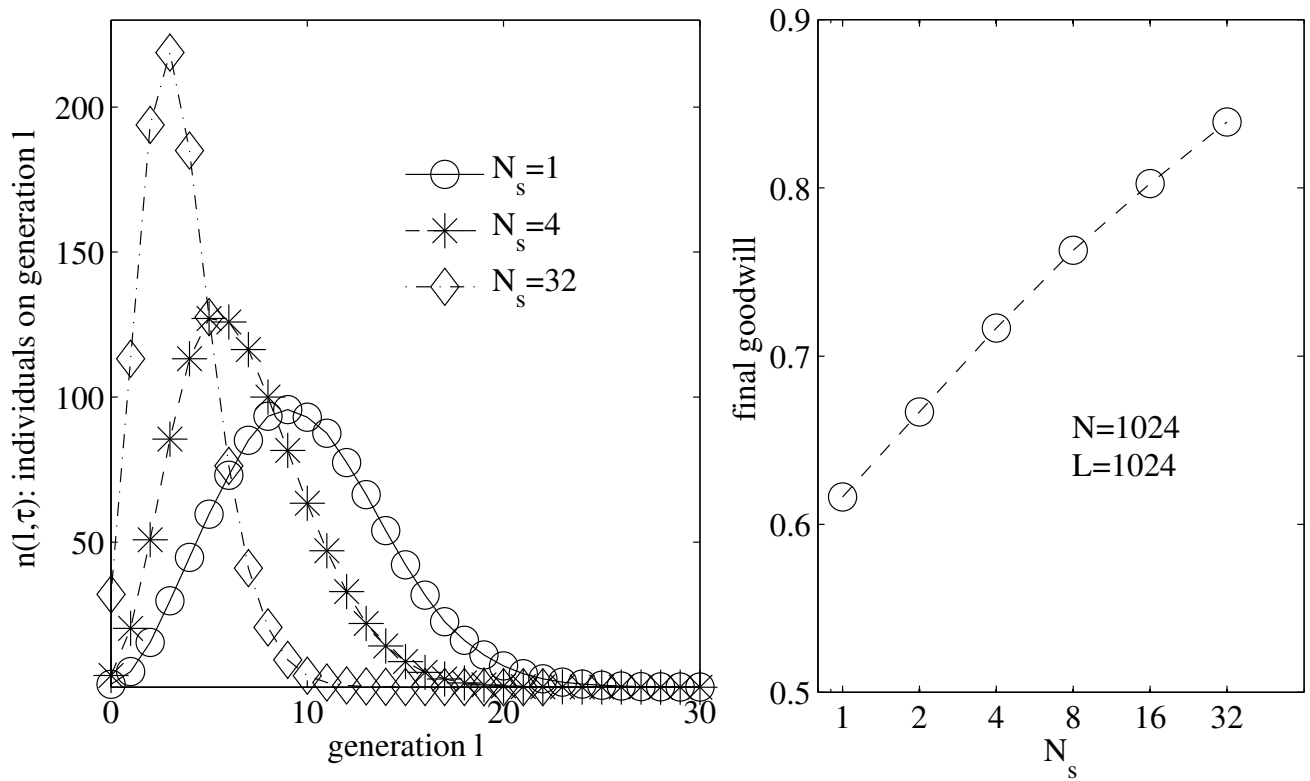

In general, a minimum value of the final goodwill corresponds to a worst-case scenario where the final number of agents that will decide to purchase the product is the least possible. Since in the model this final number is determined by the WOM process only, we can state that when the market density lies on the minimum the pure-WOM process fails. By "failure", we mean the case when it is least profitable for the firm to rely exclusively on WOM. Rather, additional advertising or more substantial initial expenditures must be taken into account in order to improve the feedback. Since the occurrence of a local minimum in the final amount of goodwill concerns systems with $\rho \sim 1$, the quantitative analysis carried out above could be useful for very crowded urban areas, and should be corroborated by local market searches.

In the following, we will suppose that the markets we study are all situated *on the right* of the goodwill minimum, hence, that the behavior they follow is monotonic: this way, we can also make quantitative forecasts about the final amount of goodwill.

Let us now consider the dependence of the final average amount of goodwill on the number of sources $N_s$ (i.e., on the initial expenditure). This function allows the firm to find its optimal sampling policy, once its initial budget (corresponding to the number of sources) and the desired final average amount of goodwill are known. An alternative analysis will be given in the next section, where the final number of purchasers is calculated from the individual goodwill instead of its average.

For any choice of $N$ and $L$, the function $\mathcal{G}_{av}$ increases monotonically with $N_s$. In Figure 4 we show how the final distribution of the population on generations and the final goodwill depend on $N_s$. On the left we show the three distributions for the values $N_s = 1, 4, 32$ (with $N = 1024$ and $L = 1024$ fixed). The distributions show a monotonic behavior, becoming more spread and peaked on higher values of $l$ as $N_s$ increases. On the right, the corresponding final goodwill is shown for those and several other values of $N_s$; the final goodwill is monotonically increasing.

The functional dependence of the final goodwill on $N_s$ depends in turn on the decay constant $z$: both the numerical fitting and the analytic approximation carried out in the Appendix show that a power law holds:

$$\mathcal{G}_{av}(N_s) = \mathcal{G}_{av}(1)N_s^{1-z}, \tag{3}$$

where we indicated with $\mathcal{G}_{av}(1)$ the final average goodwill with one initial source, and $\mathcal{G}_{av}(N_s)$ that with $N_s$ sources. Thus, if we compare two markets, say 1 and 2, with the same parameters but different number of sources $N_s^{(1)}$ and $N_s^{(2)}$, the ratio between the respective average final goodwills $\mathcal{G}_{av}^{(1)}$ and $\mathcal{G}_{av}^{(2)}$ will be

$$\frac{\mathcal{G}_{av}^{(1)}}{\mathcal{G}_{av}^{(2)}} = \left(\frac{N_s^{(1)}}{N_s^{(2)}}\right)^{1-z}. \tag{4}$$

Thus, the introduction of new sources in the market (that is, a higher expenditure) is more effective the lower is the decay parameter $z$, hence, the more "decaying" is the market.

## 6. The threshold

Once aware, an individual's attitude towards the new product is crucially affected by the degree of information he has attained. Due to the high quality of the product sampled, the more details are conveyed to the individual, the more appealing is the expected performance, which means a larger goodwill $G_j$. In fact, while passing from an individual to another, the information loses not only credibility, but also persuasiveness: those details, which are essential for an individual to purchase, may be missing. Indeed, if the individual is not aware of some characteristics of the product, he may judge it is not worth buying it.

Hence, we state that an individual $j$ adopts the new product whenever his goodwill $G_j$ is large with respect to price and risk issues. We quantify this statement by assuming that individual $j$ will purchase whenever $G_j > \xi$, where $\xi$ is a threshold value, suitably introduced and representing the price and risk "hurdles" to be overcome.

In other words, we introduce a threshold for goodwill $\xi$ beyond which the expectations (based on the amount of information) are sufficient to trigger a purchase. The threshold is associated to product price, accessibility, and availability: if they are not favorable, the individual needs strong recommendation to buy it. Of course, $\xi$ is strongly connected with the decay constant and they are both market parameters, which the firm can hardly control, but that can be analyzed by means of market surveys.

In Figure 5 we show how the percentage $\mathcal{I}$ of buyers over a population depends on the decay parameter and we compare two different values for the threshold: $\xi = 0.1$ and $\xi = 0.7$; different amounts for $N_s$ are as well represented. Interestingly, $\mathcal{I}(z)$ is a step function, whose shape is strongly affected by the chosen threshold. In fact, as shown in the appendix,

$$\mathcal{I}(z) = \frac{1}{N} \sum_{l=0}^{l_f} n(l, \tau) = \frac{\Gamma(1 + l_f, \log(N/N_s))}{l_f!}, \qquad (5)$$

where $\Gamma(a, x)$ is the incomplete gamma function and $l_f = \left\lfloor 1/\log_\xi z \right\rfloor$ is the last level corresponding to a goodwill larger than $\xi$. Hence, once $N$ and $L$ are selected, $\xi$ determines the width of the steps, while their height is determined both by $\xi$ and $N_s$. Now, the width of the $n$-th step can be calculated as:

$$\xi^{1/n} - \xi^{1/n+1} = s(s^{1/n} - 1) \qquad (6)$$

where $s = \xi^{1/(n+1)}$. Notice that the derivative $\dfrac{\partial \mathcal{I}}{\partial z}$ is just the distribution of individuals on generations, properly rescaled. Due to the extremal point exhibited by such distribution, there exists an analogous $\tilde{z}$ which maximizes the derivative and then a narrow range for $z$ where large variations on $\mathcal{I}$ are obtained by slightly changes in $z$. On the other hand, there are regions (corresponding to the tails of the distribution) where increasing the decay constant in order to increase the incomes will not be advantageous.

FIGURE 5: Percentage of buyers as a function of threshold $\xi$ and decay constant $z$.

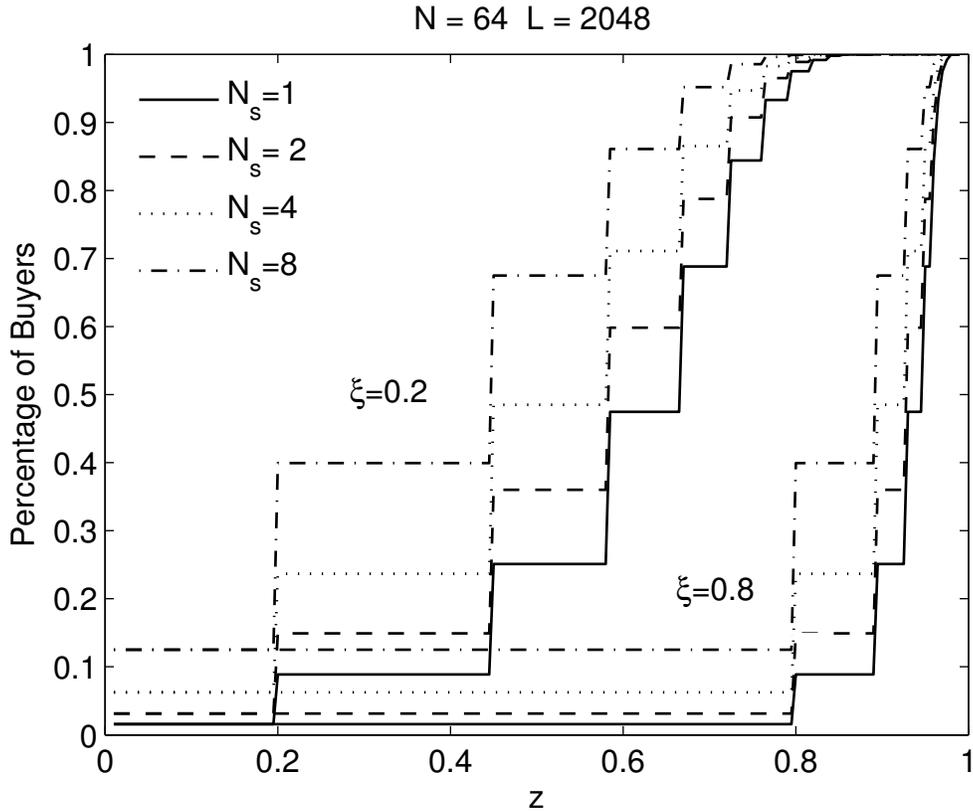

As for the role played by $N_s$, Fig. 5 shows that the percentage of buyers does not scale linearly with the initial expenditures, i.e., with the number of information sources. Nevertheless, especially when the market is characterized by a small decay parameter $z$, more initial expenditures can be employed to rise the percentage of buyers.

Finally, we notice that from Equation (5) it is possible, in principle, to estimate the value of $\xi$ *a posteriori*, namely, once the percentage of buyers and the decay constant have been properly revealed.

### 7. Summary and Conclusions

This paper introduces an innovation "micro-level" diffusion model that explicitly considers the distribution and movement of individuals in space. The main consequence of the new assumption concerns the efficiency of the word-of-mouth effect. In the particular framework we have chosen, a pre-launch stage started with a free sample distribution and without advertising, the main results are:

• A local minimum of the final goodwill for high market densities: this minimum corresponds to situations where the word-of-mouth effect is particularly weak, and advertising, or great enhancement of the firm expenditures, should be in order. It cannot be located *a priori* for a given market, but found with post-campaign market surveys, that can give indications for successive campaigns.

• An optimal sampling policy as a function of market parameters, such as the total number of potential purchasers and the geographical area covered by the market.

Future research should consider more complex situations, such as word-of-mouth effect after the launch of the product and in the presence of advertising. Still, we point out that the method we proposed here, and the qualitative results, keep holding in every model where word of mouth is considered.

### Appendix: analytical results

In the case of low density ($\rho \ll 1$) the time an informed individual walks before meeting an unaware individual becomes very large. We then assume that between each event the individuals have the time to redistribute randomly on the lattice. The number of sources is $N_s$; let $p = 5/L^2$ be the probability that two given individuals, randomly positioned on the lattice, are in contact (5 is the number of points contained in a circle of radius 1). The process is an absorbing Markov chain, with $N - N_s + 1$ states (marked with the total number of informed: $(N_s, N_s + 1, ..., N)$), one of which is an absorbing state (state $N$); the chain starts from state $N_s$. The transition matrix $\mathbf{P}$ can be written: the transition probability from state $k$ to state $n$ as a function of $N$ and $p$ is:

$$P_{k\,n} = \binom{N-k}{n-k}\left[1-\left(1-p\right)^k\right]^{n-k}\left[\left(1-p\right)^k\right]^{N-n},$$

where the indexes $k$ and $n$ run from $N_s+1$ to $N$, for any $N$ and $p$. We then make a Taylor expansion of matrix $\mathbf{P}$ to first order in $p$:

$$P_{k\,n} = \binom{N-k}{n-k}(p\,k)^{n-k}\left[1-k(N-n)\,p\right].$$

We now consider $\mathbf{Q}$, that is the submatrix built from $\mathbf{P}$ eliminating the last row and column (those corresponding to the absorbing state). The fundamental matrix $\mathbf{N} = (1-\mathbf{Q})^{-1}$ is given by:

$$N_{k\,n} = \frac{1}{n(N-n)\,p}\ \text{ for } k \le n;\ \ 0\ \text{ for } k > n.$$

The mean time $\tau$ to reach the absorbing state $N$ starting from state $N_s$ is given by the sum of the elements in the first row of $\mathbf{N}$:

$$\tau = \frac{1}{p}\sum_{n=N_s}^{N-1}\frac{1}{n(N-n)} \sim \frac{1}{N\,p}\left(\gamma + \log(\frac{N}{N_s}) + \log(N-N_s+1)\right),$$

where $\gamma = 0.577...$ is the Euler-Mascheroni constant.

The laws for $n(t)$ and $n(l,t)$ are calculated in the following way. $(1-p)^{n(t)}$ is the probability for an individual at time $t$ of not being in contact with any of the $n(t)$ informed individuals, and $\mathcal{P}(t) = 1-(1-p)^{n(t)}$ is the probability of the individual being in contact with at least one informed individual. Then the evolution of the system is governed by the equation:

$$n(t+1) = n(t) + (N-n(t))\mathcal{P}(t) = n(t) + \left(N-n(t)\right)\left(1-(1-p)^{n(t)}\right),$$

and to first order in $p$:

$$n(t+1) = n(t) + p\left(N-n(t)\right)n(t). \tag{6}$$

Since $Np = 5\rho \ll 1$, the increment of $n(t)$ at each time step is small (of order $p$), and we can take the evolution to be continuous. The equation becomes:

$$n(t+1) - n(t) \sim \frac{dn(t)}{dt} = p(N-n(t))n(t) \tag{7}$$

and the solution, with the initial condition $n(0) = N_s$, is the sigmoidal function

$$n(t) = N \frac{e^{Npt}}{e^{Npt} + N/N_s - 1}. \tag{8}$$

Let now $\mathcal{P}_l(k,s;t)$ be the probability that at time $t$ an unaware individual is in contact with $k+s$ informed individuals, of which $k$ belonging to level $l$ and $s$ belonging to some other level. Then the equation for the generation populations in the mean-field approximation is:

$$n(l,t+1) = n(l,t) + (N-n(t))\sum_{k,s}\mathcal{P}_{l-1}(k,s;t) = n(l,t) + (N-n(t))(1-(1-p)^{n(t)})\frac{n(l-1,t)}{n(t)},$$

and to first order in $p$:

$$n(l,t+1) = n(l,t) + p\,n(l-1,t)(N-n(t)).$$

Its continuous version is:

$$\frac{dn(l,t)}{dt} = p\,n(l-1,t)(N-n(t)) \tag{9}$$

that has to be solved for each $l$. For $l=1$, with the initial condition $n(1,0) = 0$, we get the solution

$$n(1,t) = N_s\Big(N\,pt - \log\big(e^{Npt}N_s + N - N_s\big) + \log N\Big) = N_s\log\!\left(\frac{n(t)}{N_s}\right).$$

We then plug this solution into Eq. (9) to get $n(2,t)$, and so on. It can be shown by induction that for every $l$, with the initial condition $n(l,t) = 0$,

$$n(l,t) = \frac{N_s^l}{l!}\Big(N\,pt - \log\big(e^{Npt}N_s + N - N_s\big) + \log N\Big)^l = \frac{N_s}{l!}\left[\log\!\left(\frac{n(t)}{N_s}\right)\right]^l. \tag{10}$$

The population distribution on generations at $t=\tau$ is

$$n(l,\tau) = \frac{N_s}{l!}\left[\log\left(\frac{N}{N_s}\right)\right]^l, \tag{11}$$

independent of $p$ (hence of $L$). This distribution has been used to fit the numerical curves.

The total information is

$$\mathcal{G}(t,z) = \sum_{l=0}^{N} n(l,t)z^l = N_s \sum_{l=0}^{N} \frac{1}{l!}\left[\log\left(\frac{n(t)}{N_s}\right)\cdot z\right]^l \sim N_s \exp\left(\log\left(\frac{n(t)}{N_s}\right)\cdot z\right) = \frac{n(t)^z}{N_s^{z-1}}. \tag{12}$$

In particular, $\mathcal{G}_{av}(\tau,z) = \mathcal{G}(\tau,z)/N = (N/N_s)^{z-1}$, in agreement with Eq. (3).

Moreover, given the threshold $\xi$, the percentage of buyers can be written as

$$\mathcal{I}(z) = \frac{1}{N}\sum_{l=0}^{l_f} n(l,\tau) = \frac{1}{N}\sum_{l=0}^{l_f} \frac{N_s}{l!}\left[\log\left(\frac{N}{N_s}\right)\right]^l = \frac{\Gamma(1+l_f,\log(N/N_s))}{l_f!}, \tag{13}$$

where we used Equation (11), which holds in limit of high dilution, and the integer $l_f = \lfloor 1/\log_\xi z \rfloor$ is the last level corresponding to a goodwill larger than the threshold: agents on levels $l > l_f$ will not buy the product.

356